\documentclass[pra,print,superscriptaddress,twocolumn]{revtex4}
\usepackage{amsfonts}
\usepackage{amssymb}
\usepackage{amsmath}
\usepackage{epsfig}
\usepackage{color}
\usepackage{graphics, graphicx}
\usepackage{bbold}
\usepackage{psfrag}
\usepackage{mathcomp}
\usepackage{subfigure}
\usepackage{verbatim}
\usepackage{color}
\usepackage[colorlinks,citecolor=blue]{hyperref}

\begin{document}

\title{Second-order topological insulator in a coinless discrete-time quantum
walk}
\author{Ya Meng}
\affiliation{State Key Laboratory of Quantum Optics and Quantum Optics Devices, Institute
of Laser spectroscopy, Shanxi University, Taiyuan 030006, China}
\author{Gang Chen}
\email{chengang971@163.com}
\affiliation{State Key Laboratory of Quantum Optics and Quantum Optics Devices, Institute
of Laser spectroscopy, Shanxi University, Taiyuan 030006, China}
\affiliation{Collaborative Innovation Center of Extreme Optics, Shanxi University,
Taiyuan, Shanxi 030006, China}
\affiliation{Collaborative Innovation Center of Light Manipulations and Applications,
Shandong Normal University, Jinan 250358, China}
\author{Suotang Jia}
\affiliation{State Key Laboratory of Quantum Optics and Quantum Optics Devices, Institute
of Laser spectroscopy, Shanxi University, Taiyuan 030006, China}
\affiliation{Collaborative Innovation Center of Extreme Optics, Shanxi University,
Taiyuan, Shanxi 030006, China}

\begin{abstract}
Higher-order topological insulators not only exhibit exotic bulk-boundary correspondence
principle, but also have an important application in quantum computing. However, they have never been achieved in quantum
walk. In this paper, we construct a two-dimensional coinless discrete-time quantum
walk to simulate second-order topological insulator with zero-dimensional
corner states. We show that both of the corner and edge states can be
observed through the probability distribution of the walker after multi-step
discrete-time quantum walks. Furthermore, we demonstrate the robustness of
the topological corner states by introducing the static disorder. Finally,
we propose a possible experimental implementation to realize this
discrete-time quantum walk in a three-dimensional integrated photonic
circuits. Our work offers a new route to explore exotic higher-order
topological matters using discrete-time quantum walks.
\end{abstract}

\maketitle

\section{Introduction}

Quantum walk, which describes the propagation of quantum particles on a
lattice \cite{QW1,QW2,QW3}, is a quantum version of classical random walk.
Due to its simplicity and high controllability, quantum walk has become a
powerful tool for universal quantum computing \cite{qwQC1,qwQC2} and
quantum simulation \cite{QS1,QS2,QS3,QS4}. Inspired by an original
theoretical paper of Kitagawa \cite{Kitagawa}, discrete-time quantum walk
(DTQW) has become an outstanding platform for simulating various topological
phenomena \cite%
{TPQWT1,TPQWT2,TPQWT3,TPQWT4,TPQWT5,TPQWT6,TPQWT7,TPQWT8,TPQWT9}. In
particular, the topological edge states and winding numbers have been detected by both unitary \cite%
{uEone1,uEone2,uEone3,uEone4,uEone5,uEone6,uEone7} and non-unitary \cite%
{nonuEone1,nonuEone2,nonuEone3,nonuEone4,nonuEone5} one-dimensional DTQWs.
For the two-dimensional case, the one-dimensional edge states have been observed without \cite%
{Etow1,Etow2} and with \cite%
{Etow3} the synthetic gauge field. Very recently, the Chern number has been successfully probed by an anomalous displacement \cite{Etow4}.

Notice that the current research on topological features of DTQWs only focus on
the simulation of the first-order topological insulator, which supports
topological protected states in the $(d-1)$-dimensional boundaries for a $d$%
-dimension system. Recently, higher-order topological insulators, which
have lower-dimensional gapless boundary states, have attracted much attention
in both theory \cite{HOT1,HOT2,HOT3,HOT4,HOT5} and experiment \cite%
{HOTe1,HOTe2,HOTe3,HOTe4,HOTe5,HOTe6}. Physically, these higher-order
topological insulators exhibit an exotic bulk-boundary correspondence
principle that an $n$th-order topological insulator supports gapless $(d-n)$%
-dimensional boundary states. Moreover, these gapless boundary states can support nontrivial fractional quasi-particles (such as parafermion or Ising anyon etc.), providing a new architecture for quantum information processing and quantum computing \cite{apply1,apply2}. Nevertheless, these interesting higher-order topological phases have never been achieved in quantum walks.

In this paper, we introduce a two-dimensional coinless DTQW to
simulate a second-order topological insulator, which hosts the
zero-dimensional corner states and one-dimensional edge states. We show that both of the corner and edge states can be
observed through the probability distribution of the walker after multi-step
DTQWs. Furthermore, we demonstrate the robustness of
the topological corner states by introducing the static disorder. Finally,
we propose a possible experimental implementation in a three-dimensional integrated photonic
circuits. Since the coupling and phase
between each two lattice sites at each step of DTQWs can be adjusted
independently, our proposal can be extended directly to realize other exotic higher-order topological insulators, such as the non-Hermitian \cite%
{noHOTI1,noHOTI2,noHOTI3} and Floquet higher-order topological insulators \cite{FHOTI1,FHOTI2,FHOTI3,FHOTI4,FHOTI5}, which have not been observed in experiments. Our work offers a new route to explore exotic topological matters using DTQWs.

This paper is organized as follows. In Sec.~\ref{section2}, we introduce a
two-dimensional coinless DTQW. In Sec.~\ref{section3}, we demonstrate the existence of
the zero-dimensional corner states and the one-dimensional edge states
through calculating the spectra and the topological invariant. In Sec.~\ref{section4}, we
show the probability distributions of the walker after multiple-step DTQWs. We also verify
the robustness of the corner states by introducing the static disorder. In
Sec.~\ref{section5}, we propose a possible experimental implementation in a
three-dimensional integrated photonic circuits. Finally, we give the
summarization in Sec.~\ref{section6}.\newline

\section{A two-dimensional coinless DTQW}\label{section2}

We begin to introduce a two-dimensional SSH model with $\pi $-flux per
plaquette, which is governed by the Hamiltonian
\begin{equation}
H=\sum_{x,y}(t_{x}a_{x+1,y}^{\dagger }a_{x,y}+t_{y}e^{ix\pi
}a_{x,y+1}^{\dagger }a_{x,y})+\text{H.c},  \label{H}
\end{equation}%
where $a_{x,y}^{\dagger }$ ($a_{x,y}$) is the creation (annihilation)
operator of a spinless particle at the site $(x,y)$, $t_{x(y)}=t+(-1)^{x(y)}%
\delta t$ are the two types of hopping amplitudes in the $x$ ($y$)
direction, respectively, and can be defined as $J_{1}=t-\delta t$, $%
J_{2}=t+\delta t$, and H.c.~is the the Hermitian conjugate. This Hamiltonian
can host the topological-protected corner states \cite{HOT1,HOT2}. In the
following, we will construct a two-dimensional coinless DTQW to simulate the
second-order topological insulator, based on the Hamiltonian (\ref{H}).

We firstly divide it into four parts
\begin{equation}
H=H_{2y}+H_{1y}+H_{2x}+H_{1x},  \label{HH}
\end{equation}%
where $H_{1x}$ ($H_{2x}$) and $H_{1y}$ ($H_{2y}$) are the intracellular
(intercellular) hoppings along the $x$ and $y$ directions, respectively.
Then, we construct a one-step operator of a DTQW as
\begin{eqnarray}
U_{\text{step}} &=&e^{-i\frac{\pi }{4}\frac{H_{2y}\Delta T}{\hbar }}e^{-i%
\frac{\pi }{4}\frac{H_{1y}\Delta T}{\hbar }}e^{-i\frac{\pi }{4}\frac{%
H_{2x}\Delta T}{\hbar }}e^{-i\frac{\pi }{4}\frac{H_{1x}\Delta T}{\hbar }}
\notag \\
&=&U_{4}U_{3}U_{2}U_{1}.  \label{U}
\end{eqnarray}%
For simplicity, we use the units $\Delta T=\hbar =1$ hereafter. For the
Hamiltonian~(\ref{H}), these four substep operators are chosen as
\begin{eqnarray}
U_{1} &=&\sum_{x=0}^{M/2-1}V_{2x+1}\otimes \mathcal{I}_{y},  \label{U1} \\
U_{2} &=&\!(|1\rangle _{x}\langle 1|+|M\rangle _{x}\langle M|)\otimes
\mathcal{I}_{y}\!+\!\!\!\sum_{x=1}^{M/2-1}V_{2x}\otimes \mathcal{I}_{y},
\label{U2} \\
U_{3} &=&\sum_{x=1}^{M}\sum_{y=0}^{M/2-1}|x\rangle \langle x|\otimes
V_{2y+1},  \label{U3} \\
U_{4} &=&\mathcal{I}_{x}\otimes (|1\rangle _{y}\langle 1|+|M\rangle
_{y}\langle M|)  \label{U4} \\
&&+\sum_{x=1}^{M}\sum_{y=1}^{M/2-1}|x\rangle \langle x|\otimes V_{2y},
\notag
\end{eqnarray}%
where the translation operators in the $x$ and $y$ directions are defined as
\begin{eqnarray}
V_{x} &=&\cos \left( \frac{\pi }{4}r\right) [|x\rangle \langle
x|+|x+1\rangle \langle x+1|]-  \label{Vx} \\
&&i\sin \left( \frac{\pi }{4}r\right) [|x+1\rangle \langle x|+|x\rangle
\langle x+1|],  \notag \\
V_{y} &=&\cos \left( \frac{\pi }{4}r\right) [|y\rangle \langle
y|+|y+1\rangle \langle y+1|]-  \label{Vy} \\
&&i\sin \left( \frac{\pi }{4}r\right) [e^{ix\pi }|y+1\rangle \langle y|+e^{%
\text{-}ix\pi }|y\rangle \langle y+1|].  \notag
\end{eqnarray}%
\begin{figure}[tbph]
\centering
\includegraphics[width=6.5cm]{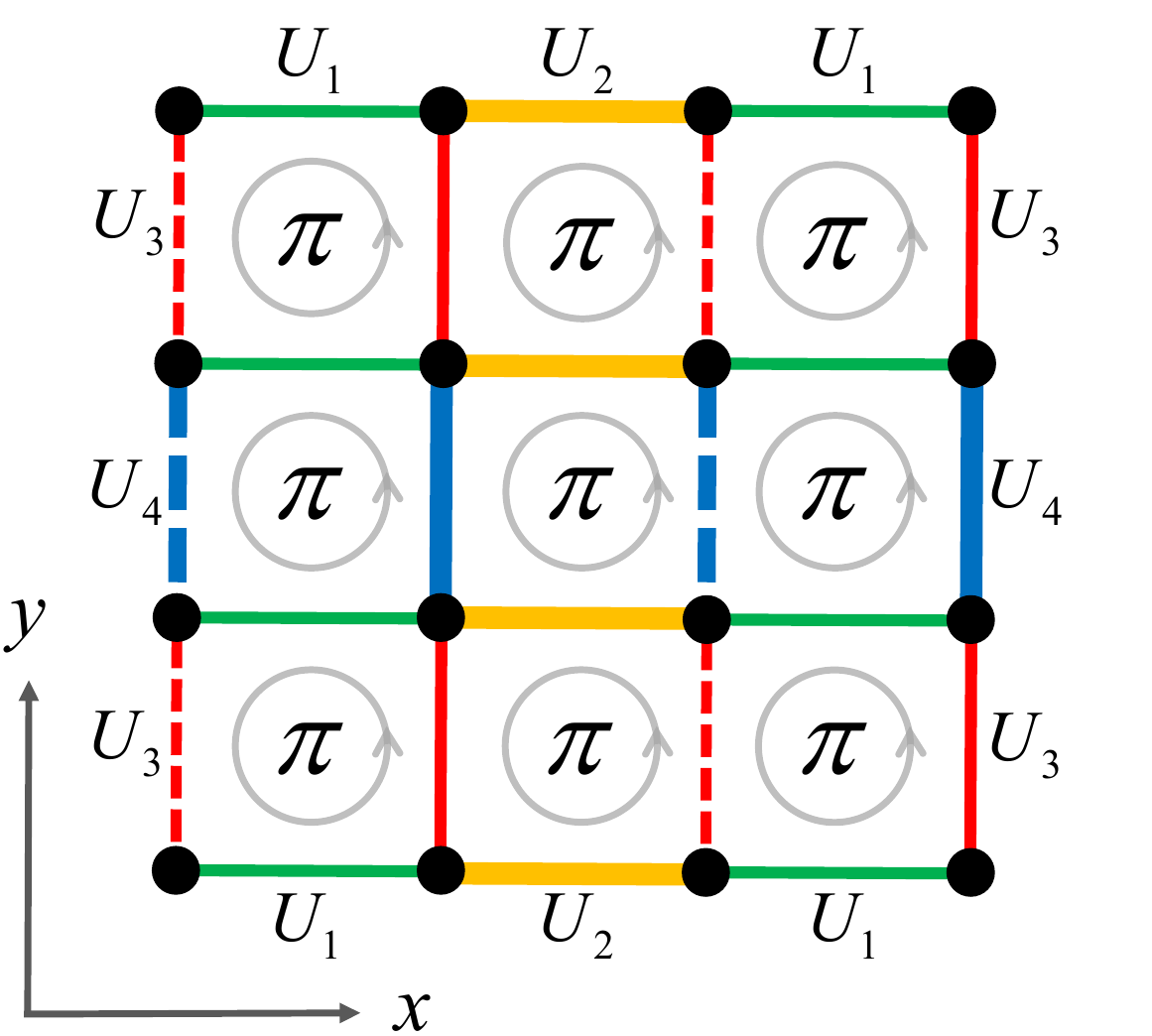}
\caption{ The proposed implementation of a coinless DTQW in a
two-dimensional lattice. The links of the lattice are marked with different
colors, degrees of thickness, and linetypes, respectively. The different
colors represent four substep operators $U_{i}$ in Eqs.~(\protect\ref{U1})-(%
\protect\ref{U4}), respectively. The links with the different thicknesses
represent two types of hopping amplitudes $J_{1}$ and $J_{2}$, respectively.
The dashed lines in the $y$ direction indicate the required $\pm\protect\pi $
phases of the hopping amplitudes, which can induce a $\protect\pi $-flux
when a walker goes through a elementary cell anticlockwise. }
\label{DTQW}
\end{figure}\newline
This DTQW is implemented in the Hilbert space $|x\rangle \otimes |y\rangle $%
, with $x\in \{1,M\}$ and $y\in \{1,M\}$ ($M$ is even). The operator $%
\mathcal{I}_{x(y)}$ denotes a $M\times M$ identity matrix in the subHilbert
space $\left\vert x\right\rangle $ ($\left\vert y\right\rangle $). It should
be emphasized that in order to generate the $\pi $-flux per plaquette, here
we have added key phase factors of the translation operator $V_{y}$. By
applying the one-step operator in Eq.~(\ref{U}) many times, a multi-step
DTQW can be realized, as shown schematically in Fig.~{\ref{DTQW}}, and the
topological-protected corner states can be explored, as will be shown.\newline

\section{Spectra and topological invariant}\label{section3}

\begin{figure}[tbph]
\centering
\includegraphics[width=8cm]{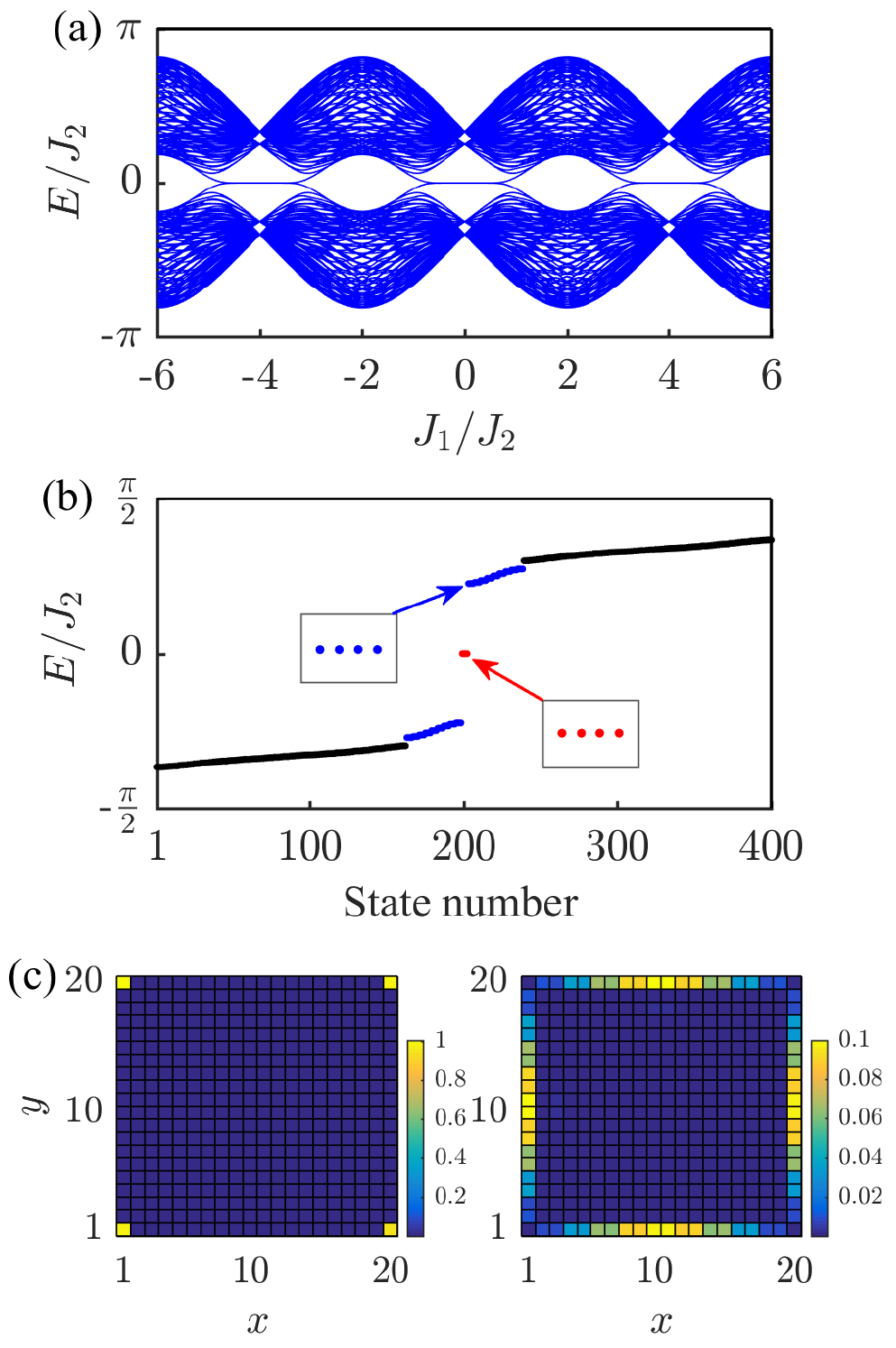}
\caption{(a) The quasienergy spectra of the effective Hamiltonian, $H_{\text{%
eff}}=i\ln U_{\text{step}}$, as a function of the parameter $J_{1}$. (b) The
quasienergy spectrum with $J_{1}=0.1$. There are four-degenerate zero- and
nonzero-energy states, denoted respectively by the red and blue points. (c)
Left: the collective distribution of the four-degenerate zero-energy
corner-localized states. Right: the collective distribution of the
four-degenerate nonzero-energy edge-localized states with the state numbers $%
\{203,204,205,206\}$. Here the lattice size is chosen as $20\times 20$ and
the parameter $J_{2}=1$.}
\label{Energy}
\end{figure}
In order to illustrate the topological features of this DTQW, here we
discuss the spectra and the topological invariant. In Fig.~\ref{Energy}(a),
we plot the quasienergy spectra of the effective Hamiltonian, $H_{\text{eff}%
}=i\ln U_{\text{step}}$, under the open boundary condition. This figure
shows clearly that the gapless zero-energy and gapped nonzero-energy states
can occur. Moreover, the gapless zero-energy states are four-degenerate and
separated from the bulk states by a large energy gap, while these
four-degenerate gapped nonzero-energy states are separated from the bulk
states only with a tiny gap, as shown in Fig.~\ref{Energy}(b). When we
increase the parameter $J_{1}$, this tiny bandgap disappears quickly. In
Fig.~\ref{Energy}(c), we plot the collective distributions of these
four-degenerate zero- and nonzero-energy states, which are indeed localized
at the four corners and edges of the lattice, respectively.

The appearance of the zero-dimensional corner states can be attributed to
the second-order bulk topology, which is described by introducing the
Wannier bands and the nested Wilson loops \cite{HOT1,HOT2}. Generally
speaking, the complete characterization of the second-order topology for a
Floquet system gives a pair of $\mathbb{Z}_{2}$ invariant, which can predict
the appearance of zero- and $\pi $-corner states \cite{FHOTI1,FHOTI2,FHOTI3,FHOTI4,FHOTI5}. For
our model, only one $\mathbb{Z}_{2}$ invariant is enough with the absence of
the $\pi$-corner states.

To construct the topological invariant, we consider the eigenstates of the
one-step operator in momentum representation
\begin{equation}
U_{\text{step}}(\mathbf{k})|E_{\mathbf{k}}\rangle =\mathrm{e}^{-iE_{\mathbf{k%
}}}|E_{\mathbf{k}}\rangle ,
\end{equation}%
where two gapped bands with the quasienergy $\pm E_{\mathbf{k}}$ are 
\begin{figure}[tbph]
\centering
\includegraphics[width=8cm]{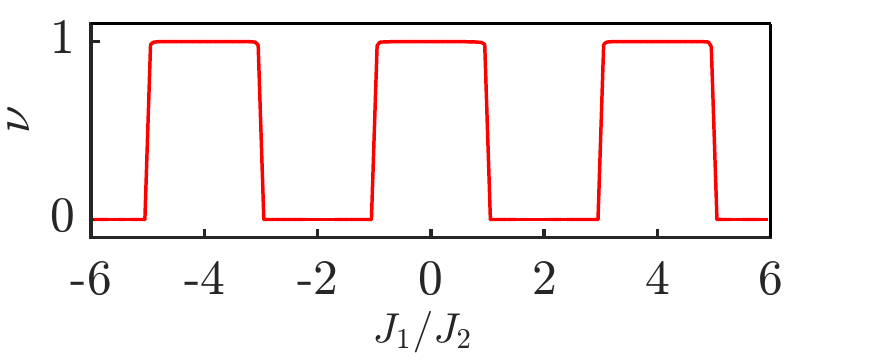}
\caption{Numerical plot of the topological invariant $\protect\nu$ by the
nested Wilson loops. The 2D Brillouin zone is discretized by using $51$ $%
\mathrm{k}$-points in each direction.}
\label{TI}
\end{figure}\newline
doubly degenerate, respectively, and the eigenstates can be denoted as $|+E_{%
\mathbf{k}}^{1}\rangle $ and $|+E_{\mathbf{k}}^{2}\rangle $ ($|-E_{\mathbf{k}%
}^{1}\rangle $ and $|-E_{\mathbf{k}}^{2}\rangle $) for upper (lower) bands.
When the lower two bands are filled, the Wilson loop operator in the $x$
direction is defined as
\begin{equation}
\mathcal{W}_{x,\mathbf{k}}=F_{x,\mathbf{k}+(N_{x}-1)\Delta k_{x}\mathbf{e}%
_{x}}\cdot \cdot \cdot F_{x,\mathbf{k}+\Delta k_{x}\mathbf{e}_{x}}F_{x,%
\mathbf{k}},  \label{Wx}
\end{equation}%
\newline
where $F_{x,\mathbf{k}}$ is a $2\times 2$ matrix with element $[F_{x,\mathbf{%
k}}]^{mn}=\langle -E_{\mathbf{k}+\Delta k_{x}\mathbf{e}_{x}}^{m}|-E_{\mathbf{%
k}}^{n}\rangle (m,n=1,2)$, $\mathbf{e}_{x}$ is the unit vector in the $x$
direction, and $\Delta k_{x}=2\pi /N_{x}$. The 2D Brillouin zone is
discretized by using the interval $(2\pi /N_{x},2\pi /N_{y})$, such that
there are $(N_{x}+1)(N_{y}+1)$ $\mathrm{k}$-points in total. With the
periodic boundary condition, $|-E_{\mathbf{k}}^{n}\rangle =|-E_{\mathbf{k}%
+2\pi \mathbf{e}_{x}}^{n}\rangle $, we diagonalize Eq.~(\ref{Wx}) as
\begin{equation}
\mathcal{W}_{x,\mathbf{k}}|\mathrm{v}_{x,\mathbf{k}}^{j}\rangle =e^{i2\pi
\mathrm{v}_{x}^{j}(k_{y})}|\mathrm{v}_{x,\mathbf{k}}^{j}\rangle ,
\end{equation}%
where $j=\pm $ denotes two Wannier bands. These Wannier bands carry their
own topological invariants, which can be evaluated by calculating the nested
Wilson loops.

\begin{figure*}[tbph]
\centering
\includegraphics[width=16cm]{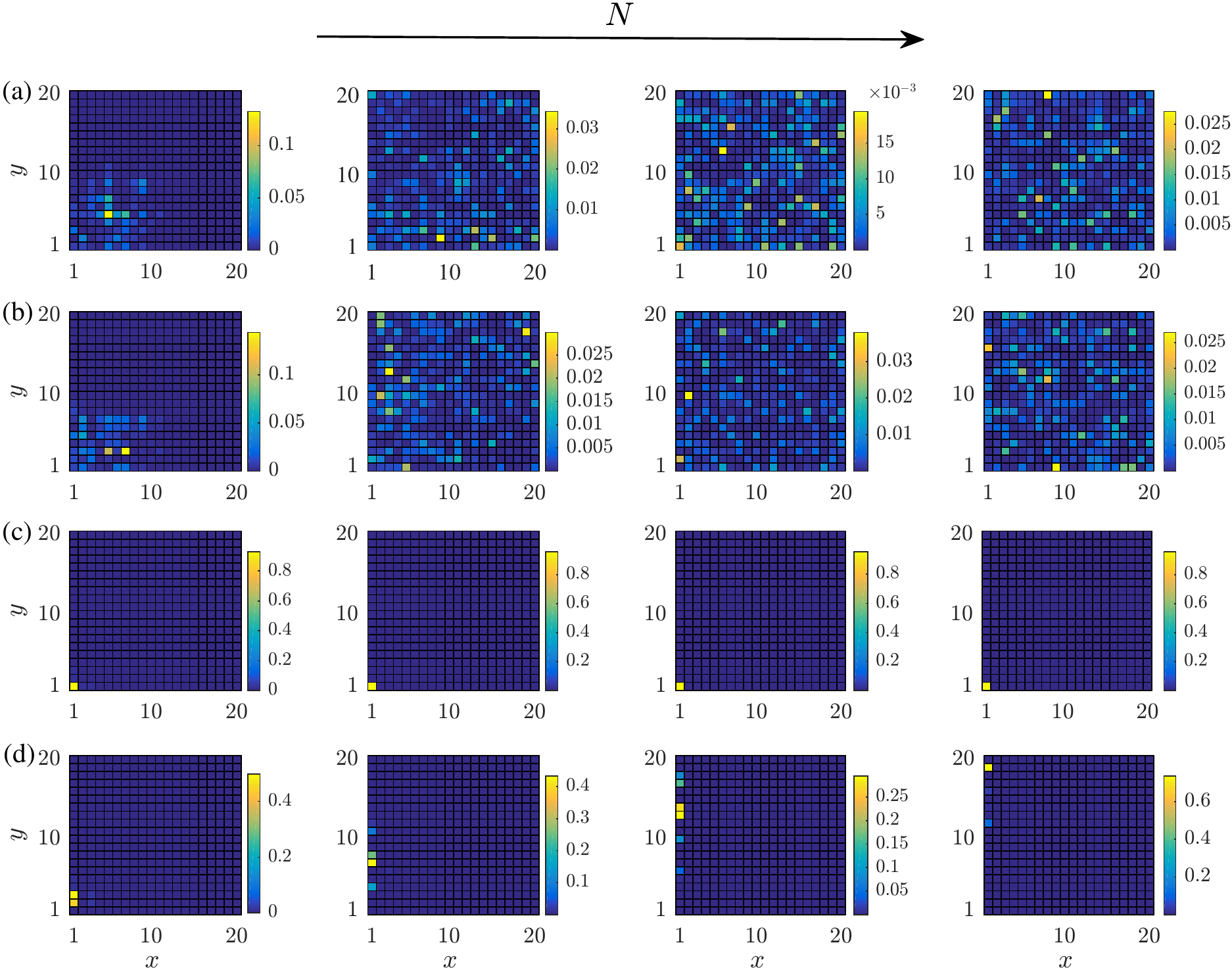}
\caption{ The probability distributions of the DTQW on a $20\times 20$
lattice for subsequent steps of $N=5,50,100,150$. The walker is initialized
at one corner $(x,y)=(1,1)$ (a,c) or edge $(x,y)=(1,2)$ (b,d) of the
lattice. In (a,b), the parameter $J_{1}=1.5$, which corresponds to a trivial
phase. In (c,d), the parameter $J_{1}=0.1$, which corresponds to a
quadrupole topological phase. }
\label{PD}
\end{figure*}
We firstly construct the Wannier states as
\begin{equation}
|w_{x,\mathbf{k}}^{\pm }\rangle =\sum\limits_{n=1,2}[\mathrm{v}_{x,\mathbf{k}%
}^{\pm }]^{n}|-E_{\mathbf{k}}^{n}\rangle ,
\end{equation}%
where $[\mathrm{v}_{x,\mathbf{k}}^{\pm }]^{n}$ denotes the $n$-th element of
the $2$-component spinor $|\mathrm{v}_{x,\mathbf{k}}^{\pm }\rangle $. Then,
with the periodic boundary condition, $|w_{x,\mathbf{k}}^{\pm }\rangle
=|w_{x,\mathbf{k}+2\pi \mathbf{e}_{y}}^{\pm }\rangle $, the nested Wilson
loops along $k_{y}$ in the Wannier bands $\mathrm{v}_{x}^{\pm }$ are
\begin{equation}
\tilde{\mathcal{W}}_{y,\mathbf{k}}=F_{y,\mathbf{k}+(N_{y}-1)\Delta k_{y}%
\mathbf{e}_{y}}^{\pm }\cdot \cdot \cdot F_{y,\mathbf{k}+\Delta k_{y}\mathbf{e%
}_{y}}^{\pm }F_{y,\mathbf{k}}^{\pm },  \label{Wy}
\end{equation}%
where $F_{y,\mathbf{k}}^{\pm }=\langle w_{x,\mathbf{k}+\Delta k_{y}\mathbf{e}%
_{y}}^{\pm }|w_{x,\mathbf{k}}^{\pm }\rangle $, $\mathbf{e}_{y}$ is the unit
vector in the $y$ direction, and $\Delta k_{y}=2\pi /N_{y}$. Through Eq.~(%
\ref{Wy}), we obtain the nested polarization as
\begin{equation}
p_{y}^{\mathrm{v}_{x}^{\pm }}=-\frac{i}{2\pi }\frac{1}{N_{x}}%
\sum\limits_{k_{x}}Log[\tilde{\mathcal{W}}_{y,\mathbf{k}}^{\pm }].
\end{equation}%
Similarly, we can also obtain the nested polarization $p_{x}^{\mathrm{v}%
_{y}^{\pm }}$ from the nest Wilson loops in the $y$ direction. Finally, the
topological quadrupole phase is characterized by a $\mathbb{Z}_{2}$
invariant \cite{HOT1},
\begin{equation}
\nu =4p_{y}^{\mathrm{v}_{x}^{\pm }}p_{x}^{\mathrm{v}_{y}^{\pm }}.
\end{equation}%
By choosing $N_{x}=N_{y}=50$, we numerically calculate the topological
invariant $\nu $ by using the above procedure. As shown in Fig.~\ref{TI}, we
find that the topological invariant $\nu $ is quantized to be $0$ or $1$,
which corresponds to the trivial or topological phases, respectively.\newline

\begin{figure}[tbph]
\centering
\includegraphics[width=7cm]{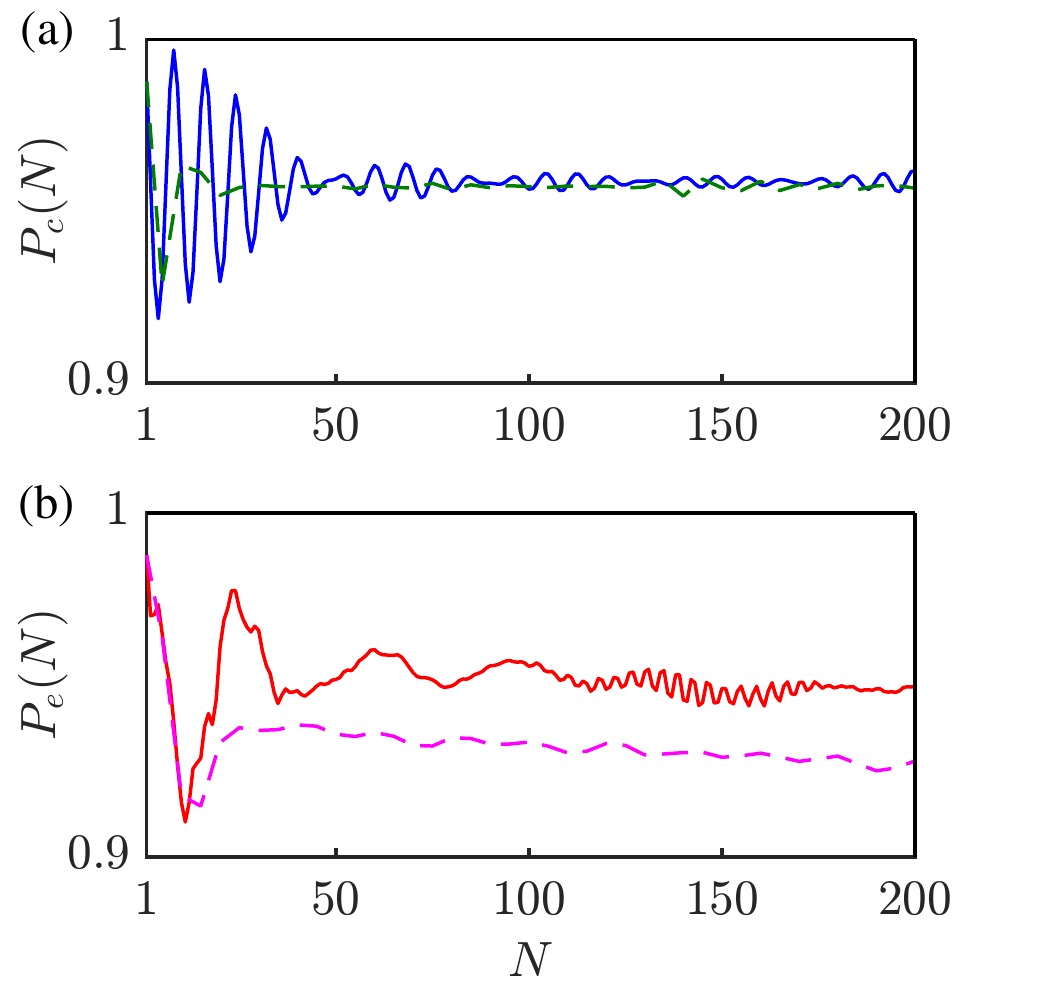}
\caption{(a) The probability $P_{c}(N)$ at the corner $(x,y)=(1,1)$ when the
walker is initialized at the same corner. (b) The probability $P_{e}(N)$ at
the right edge of the lattice when the walker is initialized at $(x,y)=(1,2)$%
. The parameter $J_{1}=0.1$. The solid lines represent the case without the
static disorder and the dashed lines represent the case $W=2.5$ averaged
over the $100$ disordered realizations.}
\label{Pt}
\end{figure}
\section{Observation of corner and edge states}\label{section4}

In this section, we mainly show that the corner and edge states can be
observed experimentally through the probability distribution of the walker
after a multi-step DTQW. It is well known that the probability distributions
of multi-step DTQWs exhibit the ballistic behaviors \cite{QW1}, which are
entirely different from the diffusive behaviors of the classic version.
Utilizing this feature, we can demonstrate the existence of the corner and
edge states through the local behavior of the probability distribution of
the walker.

In the first case, we tune the parameter $J_{1}=1.5$, which corresponds to a
trivial phase. We initialize the walker at one corner of the lattice $%
(x,y)=(1,1)$ or one edge of the lattice with $(x,y)=(1,2)$. Since the system
does not support any localized states, the probability of the walker spreads
ballistically into the bulk as increasing the number of the DTQW steps; see
Figs.~\ref{PD}(a) and~\ref{PD}(b). Then we tune the parameter to $J_{1}=$ $%
0.1$, which corresponds to a quadrupole topological phase supporting the
localized corner and edge states; see Fig.~\ref{Energy}(c). In such case,
when this initial state is prepared at one corner of the lattice $%
(x,y)=(1,1) $, since it has a large overlap with the corner state, the most
part of the walker's wave packet remains localized near $(x,y)=(1,1)$ as
increasing the step of the DTQW; see Fig.~\ref{PD}(c). In Fig.~\ref{PD}(d),
we initialize the walker at one edge of the lattice with $(x,y)=(1,2)$.
Similarly, this initial state has a large overlap with the edge states, and
we can also observe a large nonvanishing localization at one edge of the
lattice. Since the edge states have a vanishing distribution at the corners
of the lattice [see Fig.~\ref{Energy}(c)], the walker only localizes at one edge
of the lattice.

The observable properties of the corner states are robust against small
fluctuations with the second-order topological protection. Conversely, the
observable properties of the edge states are not robust due to the tiny gap
from the bulk states. To support this claim, we add the static disorder into
the evolution processing. The one-step operator for the static disorder is
introduced as
\begin{equation}
U_{\text{total}}=U_{\text{step}}\times U_{\text{dis}},
\end{equation}%
with
\begin{equation}
U_{\text{dis}}=\sum_{x,y}e^{i\delta _{x,y}}|x,y\rangle \langle x,y|.
\end{equation}%
Here $\delta _{x,y}$ is chosen randomly from the interval $[-W/2,W/2]$,
where $W$ is the disorder strength.

\begin{figure}[tbph]
\centering
\includegraphics[width=8cm]{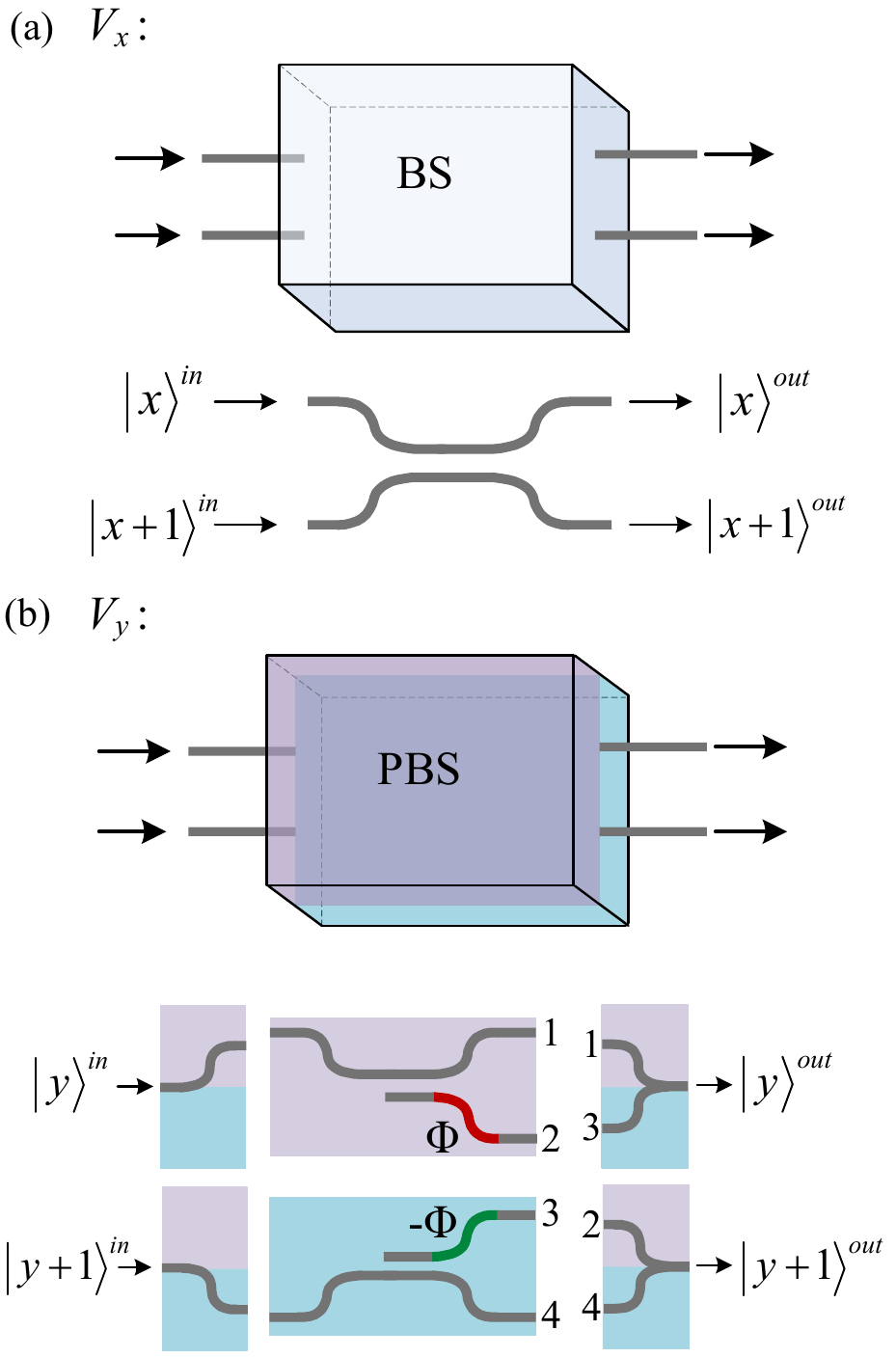}
\caption{Scheme of waveguide structure for realizing the translation
operators $V_{x}$ and $V_{y}$. (a) A single-layer waveguide structure for
realizing a beam splitter (BS). (b) A double-layer waveguide structure for
realizing a phase-shifted beam splitter (PBS).}
\label{p1}
\end{figure}
Figure~\ref{Pt} shows the probabilities $P_{c}(N)$ and $P_{e}(N)$ of the
walker remained respectively at the corner sate or the edge state without
and with the static disorder. When the walker is initialized at one corner
of the lattice $(x,y)=(1,1)$, it has a stable large probability at this
corner as increasing the step of the DTQW, and this corner probability $%
P_{c}(N)$ is robust against the static disorder; see Fig.~\ref{Pt}(a). When
the walker is initialized at one edge of the lattice $(x,y)=(1,2)$, it also
has a large probability at this edge. In Fig.~\ref{Pt}(b), we also show that
the edge probability $P_{e}(N)$ decreases when the static disorder is added,
demonstrating that the edge states are not robust. This decrease is more
evident as we increase the parameter $J_{1}$. The results for other three
corners or edges of the lattice are similar and thus not shown here.\newline

\section{Possible experimental implementation}\label{section5}

Finally, we propose a possible scheme to realize this DTQW of Eq.~(\ref{U})
in a three-dimensional integrated photonic circuit \cite{3DWG}, where a
single photon acts as a walker and a single waveguide can indicate a
two-dimensional lattice site in the $x$ and $y$ directions. The waveguides
are extended in the $z$ direction, corresponding to the time dimension of
the DTQW. The key to realize this DTQW in experiments is how to achieve the
specific translation operators $V_{x}$ and $V_{y}$ in Eqs.~(\ref{Vx})-(\ref%
{Vy}), which correspond to a beam-splitter matrix and a phased-shifted
beam-splitter matrix, respectively. Fortunately, we can realize these
translation operators in an integrated photonic circuit with the directional
coupler geometry \cite{WGDC}, where two waveguides are brought close
together for a certain interaction length and coupled by an evanescent
field. In the following, we will show how to implement the translation
operators $V_{x}$ and $V_{y}$ with the single- and double-layer waveguide
structures, respectively.

The translation operator $V_{x}$ can be realized by a directional coupler;
see Fig.~\ref{p1}(a). The standard coupled mode theory \cite{CMT} gives a
transfer matrix as%
\begin{equation}
T_{1}(z)=%
\begin{pmatrix}
\cos (Kz) & -i\sin (Kz) \\
-i\sin (Kz) & \cos (Kz)%
\end{pmatrix}%
,  \label{Tx}
\end{equation}%
which can be used to realize Eq.~(\ref{Vx}). According to Eq.~(\ref{Tx}),
the parameter in $V_{x}$ can be adjusted through altering the coupling
coefficient $K$ and the interaction length $z$.

Due to the current technology of full phase-shift controllability between
two waveguides \cite{PWG1,PWG2}, we can introduce an arbitrary phase in the
first (or second) row of $T_{1}(z)$. However, the phases required in $V_{y}$
are at the off-diagonal elements of the matrix, which indicates that we can
not realize the translation operator $V_{y}$ directly by a single
directional coupler. Thus, we design a double-layer waveguide structure to
overcome this limitation. As shown in Fig.~\ref{p1}(b), if a single photon
pulse is input from the port labeled by $\left\vert y\right\rangle ^{in}$
(or $\left\vert y+1\right\rangle ^{in}$), it will go through the upper (or
lower) layer waveguide structure and obtain a phase $\Phi $ (or $-\Phi $).
According to the coupled mode theory, the total transfer matrix governed by
this double-layer waveguide structure is
\begin{equation}
T_{2}(z)=%
\begin{pmatrix}
\cos (Kz) & -i\sin (Kz)e^{i\Phi } \\
-i\sin (Kz)e^{-i\Phi } & \cos (Kz)%
\end{pmatrix}%
,  \label{Ty}
\end{equation}%
which is exactly the phased-shifted beam-splitter matrix in Eq.~(\ref{Vy}).
Thus, the experimental implementation of $U_{\text{step}}$ is possible
with the current technology of the three-dimensional waveguide architecture
\cite{3DWG1,3DWG2,3DWG3,3DWG4}. The phase $\Phi $ can be chosen arbitrarily
and is here taken as $\Phi =m\pi $, where $m$ is an integer. When $m$ is
even, the transfer matrix $T_{2}(z)$ reduces to $T_{1}(z)$. That is, a
single-layer waveguide structure is enough for this case.\newline

\section{Conclusions}\label{section6}

In summary, we have constructed a two-dimensional coinless DTQW to simulate
the second-order topological insulator. We have shown that both of the
corner and edge states can be observed through the probability distribution
independently. Furthermore, we have demonstrated the robustness of
the topological corner states by introducing the static disorder. Finally,
we have proposed a possible experimental implementation in a three-dimensional integrated photonic
circuits. Since the coupling and phase
between each two lattice sites at each step of DTQWs can be adjusted
independently, our scheme can be generalized directly to
realize the non-Hermitian \cite%
{noHOTI1,noHOTI2,noHOTI3} and Floquet higher-order topological insulators \cite{FHOTI1,FHOTI2,FHOTI3,FHOTI4,FHOTI5}. Our work offers a new route to
explore exotic higher-order topological matters using DTQWs.\newline

\section{Acknowledgments}

This work is supported partly by the National Key R\&D Program of China
under Grant No.~2017YFA0304203; the NSFC under Grant No.~11674200; and
1331KSC.


\begin{thebibliography}{99}
\bibitem{QW1} Y. Aharonov, L. Davidovich, and N. Zagury, Quantum random
walks, Phys. Rev. A \textbf{48}, 1687 (1993).

\bibitem{QW2} E. Farhi and S. Gutmann, Quantum computation and decision trees,
Phys. Rev. A \textbf{58}, 915 (1998).

\bibitem{QW3} J. Kempe, Quantum random walks: An introductory overview,
Contemp. Phys. \textbf{44}, 307327 (2003).

\bibitem{qwQC1} A. M. Childs, Universal computation by quantum walk, Phys.
Rev. Lett. \textbf{102}, 180501 (2009).

\bibitem{qwQC2} A. M. Childs, D. Gosset, and Z. Webb, Universal computation
by multiparticle quantum walk, Science \textbf{339}, 791 (2013).

\bibitem{QS1} A. Aspuruguzik and P. Walther, Photonic quantum simulators,
Nat. Phys. \textbf{8}, 285 (2012).

\bibitem{QS2} F. Cardano, F. Massa, H. Qassim, E. Karimi, S. Slussarenko, D.
Paparo, C. de Lisio, F. Sciarrino, E. Santamato, R. W. Boyd, and L.
Marrucci, Quantum walks and wavepacket dynamics on a lattice with twisted
photons, Sci. Adv. \textbf{1}, e1500087 (2015).

\bibitem{QS3} O. Boada, L. Novo, F. Sciarrino, and Y. Omar, Quantum walks in
synthetic gauge fields with three-dimensional integrated photonics, Phys.
Rev. A \textbf{95}, 013830 (2017).

\bibitem{QS4} F. Nejadsattari, Y. Zhanf, F. Bouchard, H. Larocque, A. Sit,
E. Cohen, R. Fickler, and E. Karimi, Experimental realization of wave-packet
dynamics in cyclic quantum walks, Optica \textbf{6}, 174 (2019).

\bibitem{Kitagawa} T. Kitagawa, M. S. Rudner, E. Berg, and E. Demler,
Exploring topological phases with quantum walks, Phys. Rev. A \textbf{82},
033429 (2010).

\bibitem{TPQWT1} J. K. Asb\'{o}th, Symmetries, topological phases, and bound
states in the one-dimensional quantum walk, Phys. Rev. B \textbf{86}, 195414
(2012).

\bibitem{TPQWT2} J. K. Asb\'{o}th and H. Obuse, Bulk-boundary correspondence
for chiral symmetric quantum walks, Phys. Rev. B \textbf{88}, 121406 (2013).

\bibitem{TPQWT3} H. Obuse, J. K. Asb\'{o}th, Y. Nishimura, and N. Kawakami,
Unveiling hidden topological phases of a one-dimensional Hadamard quantum
walk, Phys. Rev. B \textbf{92}, 045424 (2015).

\bibitem{TPQWT4} J. M. Edge and J. K. Asb\'{o}th, Localization,
delocalization, and topological transitions in disordered two-dimensional
quantum walks, Phys. Rev. B \textbf{91}, 104202 (2015).

\bibitem{TPQWT5} T. Rakovszky and J. K. Asb\'{o}th, Localization,
delocalization, and topological phase transitions in the one-dimensional
split-step quantum walk, Phys. Rev. A \textbf{92}, 052311 (2015).

\bibitem{TPQWT6} T. Groh, S. Brakhane, W. Alt, D. Meschede, J. K. Asb\'{o}%
th, and A. Alberti, Robustness of topologically protected edge states in
quantum walk experiments with neutral atoms, Phys. Rev. A \textbf{94},
013620 (2016).

\bibitem{TPQWT7} T. Rakovszky, J. K. Asb\'{o}th, and A. Alberti, Detecting
topological invariants in chiral symmetric insulators via losses, Phys. Rev.
B \textbf{95}, 201407 (2017).

\bibitem{TPQWT8} V. V. Ramasesh, E. Flurin, M. Rudner, I. Siddiqi, and N. Y.
Yao, Direct probe of topological invariants using Bloch oscillating quantum
walks, Phys. Rev. Lett. \textbf{118}, 130501 (2017).

\bibitem{TPQWT9} M. Sajid, J. K. Asb\'{o}th, D. Meschede, R. F. Werner, and
A. Alberti, Creating anomalous Floquet Chern insulators with magnetic
quantum walks, Phys. Rev. B \textbf{99}, 214303 (2019).

\bibitem{uEone1} T. Kitagawa, M. A. Broome, A. Fedrizzi, M. S. Rudner, E.
Berg, I. Kassal, A. Aspuru-Guzik, E. Demler, and A. G. White, Observation of
topologically protected bound states in photonic quantum walks, Nat. Commun.
\textbf{3}, 882 (2012).

\bibitem{uEone2} F. Cardano, M. Maffei, F. Massa, B. Piccirillo, C. de
Lisio, G. De Filippis, V. Cataudella, E. Santamato, and L. Marrucci,
Statistical moments of quantum-walk dynamicas reveal topological quantum
transitions, Nat. Commun. \textbf{7}, 11439 (2016).

\bibitem{uEone3} F. Cardano, A. D'Errico, A. Dauphin, M. Maffei, B.
Piccirillo, C. D. Lisio, G. D. Filippis, V. Cataudella, E. Santamato, L.
Marrucci, M. Lewenstein, and P. Massignan, Detection of Zak phases and
topological invariants in a chiral quantum walk of twisted photons, Nat.
Commun. \textbf{8}, 15516 (2017).

\bibitem{uEone4} E. Flurin, V. V. Ramasesh, S. Hacohen-Gourgy, L. S. Martin,
N. Y. Yao, and I. Siddiqi, Observing topological invariants using quantum
walks in superconducting circuits, Phys. Rev. X \textbf{7}, 031023 (2017).

\bibitem{uEone5} X.-Y. Xu, Q.-Q. Wang, W.-W. Pan, K. Sun, J.-S Xu, G. Chen,
J.-S. Tang, M. Gong, Y.-J. Han, C.-F. Li, and G.-C. Guo, Measuring the
winding number in a large-scale chiral quantum walk, Phys. Rev. Lett.
\textbf{120}, 260501 (2018).

\bibitem{uEone6} H. Chalabi, S. Barik, S. Mittal, T. E. Murphy, M. Hafezi,
and E. Waks, Synthetic gauge field for two-dimensional time-multiplexed
quantum random walks, Phys. Rev. Lett. \textbf{123}, 150503 (2019).

\bibitem{uEone7} D.-Z. Xie, T.-S. Deng, T. Xiao, W. Gou, T. Chen, W. Yi, and
B. Yan, Topological quantum walks in momentum space with a Bose-Einstein
condensate, Phys. Rev. Lett. \textbf{124}, 050502 (2020).

\bibitem{nonuEone1} L. Xiao, X. Zhan, Z.-H. Bian, K.-K. Wang, X. Zhang,
X.-P. Wang, J. Li, K. Mochizuki, D. Kim, N. Kawakami, W. Yi, H. Obuse, B. C.
Sanders, and P. Xue, Observation of topological edge states in
parity-time-symmetric quantum walks, Nat. Phys. \textbf{13}, 1117 (2017).

\bibitem{nonuEone2} X. Zhan, L. Xiao, Z.-H. Bian, K.-K. Wang, X.-Z. Qiu, B.
C. Sanders, W. Yi, and P. Xue, Detecting topological invariants in
nonunitary discrete-time quantum walks, Phys. Rev. Lett. \textbf{119},
130501 (2017).

\bibitem{nonuEone3} K.-K. Wang, X.-Z. Qiu, L. Xiao, X. Zhan, Z.-H. Bian, B.
C. Sanders, W. Yi, and P. Xue, Observation of emergent momentum-time
skyrmions in parity-time-symmetric non-unitary quench dynamics, Nat. Commun.
\textbf{10}, 2293 (2019).

\bibitem{nonuEone4} K.-K. Wang, X.-Z. Qiu, L. Xiao, X. Zhan, Z.-H. Bian, W.
Yi, and P. Xue, Simulating dynamic quantum phase transitions in photonic
quantum walks, Phys. Rev. Lett. \textbf{122}, 020501 (2019).

\bibitem{nonuEone5} L. Xiao, K.-K. Wang, X. Zhan, Z.-H. Bian, K. Kawabata,
M. Ueda, W. Yi, and P. Xue, Observation of critical phenomena in
parity-time-symmetric quantum dynamics, Phys. Rev. Lett. \textbf{123},
230401 (2019).

\bibitem{Etow1} B. Wang, T. Chen, and X. Zhang, Experimental observation of
topologically protected bound states with vanishing Chern numbers in a
two-dimensional quantum walk, Phys. Rev. Lett. \textbf{121}, 100501 (2018).

\bibitem{Etow2} C. Chen, X. Ding, J. Qin, Y. He, Y. Luo, M. Chen, C. Liu, X.
Wang, W. Zhang, H. Li, L. You, Z. Wang, D. Wang, B. C. Sanders, C. Lu, and
J. Pan, Observation of topologically protected edge states in a photonic
two-dimensional quantum walk, Phys. Rev. Lett. \textbf{121}, 100502 (2018).

\bibitem{Etow3} H. Chalabi, S. Barik, S. Mittal, T. E. Murphy, M. Hafezi,
and E. Waks, Synthetic gauge field for two-dimensional time-multiplexed
quantum random walks, Phys. Rev. Lett. \textbf{123}, 150503 (2019).

\bibitem{Etow4} A. D'Errico, F. Cardano, M. Maffei, A. Dauphin, R. Barboza,
C. Esposito, B. Piccirillo, M. Lewenstein, P. Massignan, and L. Marrucci,
Two-dimensional topological quantum walks in the momentum space of
structured light, Optica \textbf{7}, 108 (2020).

\bibitem{HOT1} W. A. Benalcazar, B. A. Bernevig, and T. L. Hughes, Quantized
electric multipole insulators, Science \textbf{357}, 61 (2017).

\bibitem{HOT2} W. A. Benalcazar, B. A. Bernevig, and T. L. Hughes, Electric
multiple moments, topological multipolemoment pumping, and chiral hinge
states in crystalline insulators, Phys. Rev. B \textbf{96}, 245115 (2017).

\bibitem{HOT3} J. Langbehn, Y. Peng, L. Trifunovic, F. von Oppen, and P. W.
Brouwer, Reflection-symmetric second-order topological insulators and
superconductors, Phys. Rev. Lett. \textbf{119}, 246401 (2017).

\bibitem{HOT4} Z. Song, Z. Fang, and C. Fang, (d-2)-dimensional edge states
of rotation symmetry protected topological states, Phys. Rev. Lett. \textbf{%
119}, 246402 (2017).

\bibitem{HOT5} F. Schindler, A. M. Cook, M. G. Vergniory, Z. Wang, S. S. P.
Parkin, B. A. Bernevig, and T. Neupert, Higher-order topological insulators,
Sci. Adv. \textbf{4}, eaat0346 (2018).

\bibitem{HOTe1} M. Serra-Garcia, V. Peri, R. S\"{u}sstrunk, O. R. Bilal, T.
Larsen, L. G. Villanueva, and S. D. Huber, Observation of a phononic
quadrupole topological insulator, Nature (London) \textbf{555}, 342 (2018).

\bibitem{HOTe2} C. W. Peterson, W. A. Benalcazar, T. L. Hughes, and G. Bahl,
A quantized microwave quadrupole insulator with topologically protected
corner states, Nature (London) \textbf{555}, 346 (2018).

\bibitem{HOTe3} F. Schindler, Z.-J. Wang, M. G. Vergniory, A. M. Cook, A.
Murani, S. Sengupta, A. Y. Kasumov, R. Deblock, S. Jeon, I. Drozdov, H.
Bouchiat, S. Gu\'{e}ron, A. Yazdani, B. A. Bernevig, and T. Neupert,
Higher-order topology in bismuth, Nat. Phys. \textbf{14}, 918 (2018).

\bibitem{HOTe4} S. Imhof, C. Berger, F. Bayer, J. Brehm, L. Molenkamp, T.
Kiessling, F. Schindler, C. H. Lee, M. Greiter, T. Neupert, and R. Thomale,
Topological-circuit realization of topological corner modes, Nat. Phys.
\textbf{14}, 925 (2018).

\bibitem{HOTe5} S. Mittal, V. V. Orre, G. Zhu, M. A. Gorlach, A. Poddubny,
and M. Hafezi, Photonic quadrupole topological phases, Nat. Phys. \textbf{13}%
, 692 (2019).

\bibitem{HOTe6} F. Zangeneh-Nejad and R. Fleury, Nonlinear second-order
topological insulators, Phys. Rev. Lett. \textbf{123}, 093502 (2019).

\bibitem{apply1} Y. You, D. Litinski and F. von Oppen, Higher-order topological superconductors as generators of quantum codes, Phys. Rev. B \textbf{100}, 054513 (2019).

\bibitem{apply2} K. Laubscher, D. Loss, and J. Klinovaja, Fractional topological superconductivity and parafermion corner states, Phys. Rev. Research \textbf{1}, 032017 (2019).

\bibitem{noHOTI1} C. H. Lee, L. Li, and J. Gong, Hybrid higher-order
skin-topological modes in nonreciprocal systems, Phys. Rev. Lett. \textbf{123%
}, 016805 (2019).

\bibitem{noHOTI2} T. Liu, Y.-R. Zhang, Q. Ai, Z. Gong, K. Kawabata, M. Ueda,
and F. Nori, Second-order topological phases in non-Hermitian systems, Phys.
Rev. Lett. \textbf{122}, 076801 (2019).

\bibitem{noHOTI3} X.-W. Luo and C.-W. Zhang, Higher-order topological corner
states induced by gain and loss, Phys. Rev. Lett. \textbf{123}, 073601
(2019).

\bibitem{FHOTI1} M. Rodriguez-Vega, A. Kumar, and B. Seradjeh, Higher-order Floquet topological phases with corner and bulk
bound states, Phys. Rev. B \textbf{100}, 085138 (2019).

\bibitem{FHOTI2} R. W. Bomantara, L. Zhou, J. Pan, and J. Gong,
Coupled-wire construction of static and Floquet second-order topological insulators, Phys. Rev. B \textbf{99}, 045441
(2019).

\bibitem{FHOTI3} R. Seshadri, A. Dutta, and D. Sen, Generating a second-order topological insulator with multiple corner states by
periodic driving, Phys. Rev. B \textbf{100}, 115403 (2019).

\bibitem{FHOTI4} Y. Peng and G. Refael, Floquet second-order topological
insulators from nonsymmorphic space-time symmetries, Phys. Rev. Lett.
\textbf{123}, 016806 (2019).

\bibitem{FHOTI5} H. Hu, B. Huang, E. Zhao, and W. Vincent Liu, Dynamical
singularities of Floquet higher-order topological insulators, Phys. Rev.
Lett. \textbf{124}, 057001 (2020).

\bibitem{3DWG} S. Gross and M. J. Withford, Ultrafast-laser-inscribed 3D
integrated photonics: challenges and emerging applications, Nanophotonics
\textbf{4}, 332 (2015).

\bibitem{WGDC} F. Flamini, N. Spagnolo, and F. Sciarrino, Two-particle
bosonic-fermionic quantum walk via integrated photonics, Phys. Rev. Lett.
\textbf{108}, 010502 (2012).

\bibitem{CMT} K. Okamoto, Funfamentals and applications of optical
waveguides (Elsevier, San Diego, 2006).

\bibitem{PWG1} A. Crespi, R. Osellame, R. Ramponi, D. J. Brod, E. F. Galv%
\~{a}o, N. Spagnolo, C. Vitelli, E. Maiorino, P. Mataloni, and F. Sciarrino,
Integrated multimode interferometers with arbitrary designs for photonic
boson sampling, Nat. Photonics \textbf{7}, 545 (2013).

\bibitem{PWG2} A. Crespi, R. Osellame, R. Ramponi, V. Giovannetti, R. Fazio,
L. Sansoni, F. D. Nicola, F. Sciarrino, and P. Mataloni, Anderson
localization of entangled photons in an integrated quantum walk, Nat.
Photonics \textbf{7}, 322 (2013).

\bibitem{3DWG1} S. Nolte, M. Will, J. Burghoff, and A. Tuennermann,
Femtosecond waveguide writing: a new avenue to three-dimensional integrated
optics, Appl. Phys. A Mater. Sci. Process. \textbf{77}, 109 (2003).

\bibitem{3DWG2} N. Spagnolo, C. Vitelli, L. Sansoni, E. Maiorino, P.
Mataloni, F. Sciarrino, D. J. Brod, E. F. Galva\~{o}, A. Crespi, R. Ramponi,
and R. Osellame, General rules for bosonic bunching inmultimode
interferometers, Phys. Rev. Lett. \textbf{111}, 130503 (2013).

\bibitem{3DWG3} H. Tang, X.-F. Lin, Z. Feng, J.-Y. Chen, J Gao, K Sun, C.-Y.
Wang, P.-C. Lai, X.-Y. Xu, Y Wang, L.-F. Qiao, A.-L. Yang, and X.-M. Jin,
Experimental two-dimensional quantum walk on a photonic chip, Sci. Adv.
\textbf{4}, eaat3174 (2018).

\bibitem{3DWG4} H. Tang, C. D. Franco, Z.-Y. Shi, T.-S. He, Z. Feng, J. Gao,
K. Sun, Z.-M. Li, Z.-Q. Jiao, T-Y Wang, M. S. Kim, and X.-M. Jin,
Experimental quantum fast hitting on hexagonal graphs, Nat. Photonics
\textbf{12}, 754 (2018).

\end{thebibliography}
\end{document}